\documentclass{pasj00}

\begin{document}
\SetRunningHead{M. Uemura et al.}{Active Phase of V4641 Sgr in 2002}
\Received{2002//}
\Accepted{2002//}

\title{Outburst and Post-Outburst Active Phase of the Black Hole X-ray
Binary V4641 Sgr in 2002}

\author{
Makoto \textsc{Uemura}\altaffilmark{1},
Taichi \textsc{Kato}\altaffilmark{1},
Ryoko \textsc{Ishioka}\altaffilmark{1},
Kenji \textsc{Tanabe}\altaffilmark{2},
Ken'ichi \textsc{Torii}\altaffilmark{3},\\
Roland \textsc{Santallo}\altaffilmark{4},
Berto \textsc{Monard}\altaffilmark{5},
Craig B. \textsc{Markwardt}\altaffilmark{6},
Jean H. \textsc{Swank}\altaffilmark{7},\\
Robert J. \textsc{Sault}\altaffilmark{8},
Jean-Pierre \textsc{Macquart}\altaffilmark{9},
Michael \textsc{Linnolt}\altaffilmark{10},
Seiichiro \textsc{Kiyota}\altaffilmark{11},\\
Rod \textsc{Stubbings}\altaffilmark{12},
Peter \textsc{Nelson}\altaffilmark{13},
Tom \textsc{Richards}\altaffilmark{14},
Charles \textsc{Bailyn}\altaffilmark{15},
Doug \textsc{West}\altaffilmark{16},\\
Gianluca \textsc{Masi}\altaffilmark{17},
Atsushi \textsc{Miyashita}\altaffilmark{18},
Yasuo \textsc{Sano}\altaffilmark{19},
and
Toni \textsc{Scarmato}\altaffilmark{20}}
\altaffiltext{1}{Department of Astronomy, Faculty of Science, Kyoto
University, Sakyou-ku, Kyoto 606-8502}
\email{uemura@kusastro.kyoto-u.ac.jp}
\altaffiltext{2}{Department of Biosphere-Geosphere Systems, Faculty of
Informatics, \\Okayama University of Science, Ridaicho 1-1, Okayama
700-0005}
\altaffiltext{3}{Cosmic Radiation Laboratory, RIKEN (Institute of
Physical and Chemical Research), 2-1,\\ Hirosawa, Wako, Saitama 351-0198}
\altaffiltext{4}{Southern Stars Observatory IAU \# 930, Tahiti French
Polynesia}
\altaffiltext{5}{Bronberg Observatory, PO Box 11426, Tiegerpoort 0056, South
Africa}
\altaffiltext{6}{Department of Astronomy, University of Maryland,
College Park, MD 20742, USA}
\altaffiltext{7}{Laboratory for High Energy Astrophysics, NASA Goddard Space
Flight Center, Mail Code 662,\\ Greenbelt, MD 20771, USA}
\altaffiltext{8}{Australia Telescope National Facility, Narrabri, NSW
2390, Australia}
\altaffiltext{9}{Kapteyn Astronomical Institute, University of
Groningen, Postbus 800, 9700 AV Groningen,\\ The Netherlands}
\altaffiltext{10}{PO Box 11401, Honolulu, HI 96828, USA}
\altaffiltext{11}{Variable Star Observers League in Japan (VSOLJ); Center for
Balcony Astrophysics, \\1-401-810 Azuma, Tsukuba 305-0031}
\altaffiltext{12}{19 Greenland Drive, Drouin 3818, Victoria, Australia}
\altaffiltext{13}{RMB 2493, Ellinbank 3820, Australia}
\altaffiltext{14}{Woodridge Observatory, 8 Diosma Rd, Eltham, Vic 3095,
Australia}
\altaffiltext{15}{Department of Astronomy, Yale University, P.O. Box
208101, New Haven, CT 06520-8101, USA}
\altaffiltext{16}{West Skies Observatory, P.O. Box 517, Derby, KS 67037,
USA.}
\altaffiltext{17}{Via Madonna de Loco, 47, 03023 Ceccano (FR), Italy}
\altaffiltext{18}{Seikei High School, 3-10-13 Kita-machi, Kichijouji,
Musashino, Tokyo 180-8633}
\altaffiltext{19}{VSOLJ, 3-1-5 Nishi Juni-jou Minami, Nayoro, Hokkaido
096-0022}
\altaffiltext{20}{via Cuppari 10, 89817 San Costantino di Briatico (VV),
Calabria, Italy}


%


\maketitle

\begin{abstract}
 The black hole X-ray binary V4641 Sgr experienced an outburst in 2002
 May which was detected at X-ray, optical, and radio wavelengths.  The
 outburst lasted for only 6 days, but the object remained active for
 the next several months.  Here we report on the detailed properties of
 light curves during the outburst and the post-outburst active phase.
 We reveal that rapid optical variations of $\sim 100\;{\rm s}$ became
 more prominent when a thermal flare weakened and the optical spectrum
 flattened in the $I_{\rm c}$, $R_{\rm c}$, and $V$-band region.  In
 conjunction with the flat spectrum in the radio range, this strongly
 indicates that the origin of rapid variations is not thermal emission,
 but synchrotron emission.  Just after the outburst, we
 detected repeated flares at optical and X-ray wavelengths.  The optical
 and X-ray light curves exhibited a strong correlation, with the X-rays,
 lagging by about 7 min.  The X-ray lag can be understood in terms of a
 hot region propagating into the inner region of the accretion flow.
 The short X-ray lag, however, requires modifications of this simple
 scenario to account for the short propagation time.  We also detected
 rapid optical variations with surprisingly high amplitude 50 days after
 the outburst, which we call optical flashes.  During the most prominent
 optical flash, the object brightened by 1.2 mag only within 30 s.  The
 released energy indicates that the emission source should be at the
 innermost region of the accretion flow.
\end{abstract}

\section{Introduction}

Black hole X-ray binaries are close binary systems which contain a
stellar mass black hole of $\sim 10M_{\rm \solar}$ and a main sequence
or an evolved secondary star (\cite{tan95BHXN}).  The gas from the
secondary star forms an accretion disk around the black hole, which
causes various types of variations (\cite{vanderkli89QPO};
\cite{odo96BHXNSH}).  X-ray rapid variations are of interest because
their short time scales of 0.1--1000 Hz indicate their emission
originates from the inner region of the accretion flow
(e.g. \cite{mor97grs1915QPO}; \cite{wei97CygX1}). The variations
of the X-ray flux hence provide important clues concerning the physics
of the accretion flow, and on the nature of the black holes
themselves(\cite{wei98BHspin}; \cite{abr01BHspin}). 

Optical rapid variations from black hole binaries have recently received
attention as unique probes of accretion and jet physics
(\cite{kan01nature}; \cite{uem02v4641letter}).  Rapid optical variability
was first detected in the black hole binary GX 339$-$4 during
an optically very bright state ($V=15.4$) in 1981, when it showed 20\,s
optical QPOs with relatively large amplitudes, of order 0.1 mag
(\cite{mot82gx339}).  QPOs of similar amplitude but a 190\,s period were
detected during an optically moderate brightness state
(\cite{cam90gx339}).  Smaller-amplitude QPOs, of order 0.01 mag, with
periods of 7--16 s were also reported (\cite{mot85gx339};
\cite{ima90gx339}; \cite{cam97GX339}).  Another black hole binary, V404
Cyg, exhibited possible 0.7 mag, 10 min variations around the maximum of
its X-ray nova outburst (\cite{bui89v404cyg};
\cite{wag89v404cygiauc4797}).  The object again showed 3--10 min, 0.01
mag QPOs during the decline from the outburst (\cite{got91v404cyg}).
XTE J1118+480 recently provided a rare opportunity to observe rapid
variations simultaneously at optical and X-ray wavelengths
(\cite{kan01nature}).  Its 0.1 mag optical variations on a time scale of
seconds exhibited a strong correlation with X-ray variations, with a
preceding optical dip (\cite{kan01nature}; \cite{spr02j1118}).  GX
339$-$4 exhibited similar optical dips, which were anticorrelated with
the X-ray flares (\cite{mot83gx339}).  These rapid optical variations
tend to appear during the low/hard state.  On the other hand, $\lesssim
30\;{\rm s}$ variations are reported in both A0620-00 and X-ray Nova Mus
1991 even in the quiescent state (\cite{hyn03v616mon}).

The optical emission is generally considered to be thermal
emission near the outer portion of the accretion disk, where the
temperature is relatively low, of order $10^4\;{\rm K}$
(\cite{tan95BHXN}).  The short-time scale of the rapid variations
implies that the emission originates in the inner part of the accretion
flow.  The spectrum expected for an optically-thick, geometrically-thin
disk (the so called standard disk) predicts that most of the released energy
is observed in the X-ray range, and only weak optical emission is
expected from such an inner region.  The rapid optical variations should
hence originate from a non-thermal source, and the most promising
candidate for this emission mechanism is synchrotron radiation
(\cite{fab82gx339}; \cite{kan01nature}).

It has been proposed that strong synchrotron emission can significantly
contribute to the optical flux observed in black hole binaries.
\citet{fen01lowhardjet} proposes that the low/hard states of black hole
binaries are characterized by self-absorbed synchrotron emission from
jets observed at radio wavelengths.  The synchrotron emission from jets
is proposed to extend to the infrared, and possibly to the optical.  
In the case of large-scale jets, a near-infrared jet was directly imaged
in GRS 1915+105 (\cite{sam96IRjet}).  The $\sim 1000$\,s infrared
variations in this object precede the radio variations, and are widely
believed to be synchrotron flares associated with the jet
(\cite{mir98grs1915}).  Magnetic flares at the inner accretion flow are
another proposed source of strong synchrotron emission
(\cite{mer00j1118}). 

V4641 Sgr is an X-ray binary system with an orbital period of 2.8\,d,
containing a secondary star of 5--8$M_{\solar}$ and a black hole of
$\sim 9.6 M_{\solar}$ (\cite{oro01v4641sgr}).  This object first
received attention during a luminous outburst in 1999 September.
Despite its high luminosity, $\sim 10^{39}\;{\rm erg\,s^{-1}}$, the
duration of the outburst was quite short; it lasted a few hours in the
X-ray range (\cite{smi99v4641}) and one day in the optical range
(\cite{uem02v4641}).  This system also possesses highly relativistic
jets, which were detected by radio observations during this outburst
(\cite{hje00v4641}).  The source's rapid state transitions make it an
ideal candidate to study the relationship between the accretion flow and
the jet with multi-wavelength observations (\cite{mir98grs1915};
\cite{eik98grs1915}).  Activity was also reported in the radio range
in 2000,\footnote{$\langle$http://vsnet.kusastro.kyoto-u.ac.jp/vsnet/Mail/vsnet-campaign-xray/msg00009.html,
msg00011.html, msg00020.html$\rangle$} which implies that V4641 Sgr
experiences active states frequently, in common with GRS 1915+105, but
unlike most typical black hole binaries whose quiescent states last for
over ten years (\cite{rao00grs1915QPO}; \cite{tan95BHXN}).

The most recent major outburst occurred in 2002 May.  During this
outburst, we detected optical short-term modulations, a part of which was
reported in \citet{uem02v4641letter}.  Here we report the detailed
features of V4641 Sgr during the 2002 May outburst and the post-outburst
active phase. In the next section, we present a short summary of our
observations.  In section 3, we report the results of our
observations.  We then discuss the nature of the emission source during
the active phase in section 4.  The summary of this paper is presented in
the final section.

\section{Observation}

We performed CCD photometric observations at 15 optical observatories from
JD 2452397 to 2452569.  The observations were typically performed with
30-cm class telescopes.  The journal of our CCD observations is given in
table \ref{tab:log}.  We used the standard reduction method for obtained
images, and adjusted unfiltered CCD magnitude systems to the $R_{\rm
c}$-system with the same manner described in \citet{uem02v4641letter}.

The X-ray data were taken by the RXTE Porportional Counter Array (PCA)
using pointed observations from 24.15--28.65 May 2002 (UT).  The PCA
consists of five Proportional Counter Units (PCUs) which are sensitive
in the 2--60 keV X-ray band (Jahoda et al. 1996).  The fields of view of
the five PCUs are restricted to $1^\circ$ FWHM by hexagonal collimators
which are coaligned ($2.2^\circ$ full width at zero response).  Data
were filtered using the standard procedures for bright sources
recommended by the RXTE Guest Observer Facility.  The total good time
was approximately 36 ks.  Data from the Standard2 mode were used,
which have a time binning of 16 sec and moderate spectral resolution.
Counts from the top detector layer in the energy range 2--10 keV were
selected.  Particle backgrounds were subtracted using the "CMVLE"
model, and an additional estimated background of 1.0 mCrab due to
galactic diffuse emission was also subtracted.  The quoted X-ray flux
values were found by dividing the total count rate by the number of
active PCUs, and scaling to the mean PCA count rate of the Crab in the
same energy band.  The X-ray outburst consisted of a period of about 2
days from 24--26 May where the flux varied from approximately 5--50
mCrab, with large fluctuations on time scales of minutes.  After 26.0
May, no significant persistent flux was detected, however there
continued to be X-ray flares with peak fluxes as high as 36 mCrab.

Radio observations were performed with the Australia Telescope Compact
Array (ATCA) on JD~ 2452419 and briefly again on the two subsequent
days.   The ATCA is an Earth-rotation aperture synthesis array,
comprising six 22~m antennas which can be moved along an east-west track
to give baselines up to 6~km (\cite{fra92ATCA}).  The source was
observed at bands centered on frequencies of 1.384, 2.496, 4.800 and
8.640 GHz with 128\,MHz bandwidth in two orthogonal polarizations.  

\begin{longtable}{cccccc}
\caption{Observation log}
\label{tab:log}
\hline\hline
$T_{\rm start}$ (HJD) & $\Delta T$ (hr) & $T_{\rm exp}$ & N & Filter & Site\\
\hline
\endhead
\hline
{\small $T_{\rm exp}$: Exposure time, N: Number of frame}
\endfoot
\hline
2452397.281 & 0.53 & 10 & 117 & -- & Kyoto\\
2452406.264 & 0.20 & 10 & 38 & -- & Kyoto\\
2452407.297 & 0.18 & 10 & 37 & -- & Kyoto\\
2452415.398 & 6.41 & 10 & 237 & -- & Bronberg\\
2452415.667 & 6.33 & 120 & 8 & $V$ & Cerro Tololo\\
2452415.908 & 0.65 & 15 & 18 & $V$ & Mulvane\\
2452415.913 & 0.24 & 30 & 14 & $V$ & Mulvane\\
2452416.248 & 1.37 & 10 & 261 & -- & Kyoto\\
2452416.296 & 9.05 & 10 & 904 & -- & Bronberg\\
2452416.742 & 1.85 & 120 & 24 & $V$ & Cerro Tololo\\
2452416.913 & 0.62 & 30 & 35 & -- & Mulvane\\
2452417.544 & 3.00 & 10 & 948 & -- & Bronberg\\
2452417.665 & 4.94 & 120 & 23 & $V$ & Cerro Tololo\\
2452417.916 & 0.46 & 30 & 26 & -- & Mulvane\\
2452418.202 & 1.82 & 1 & 1780 & $R_{\rm c}$ & Ouda\\
2452418.208 & 1.30 & 5 & 454 & -- & Okayama\\
2452418.246 & 0.69 & 10 & 115 & -- & Kyoto\\
2452418.383 & 2.11 & 10 & 293 & -- & Bronberg\\
2452419.022 & 2.14 & 15 & 203 & $V$ & Ellinbank\\
2452419.086 & 4.85 & 30 & 137 & $I_{\rm c}$ & Tsukuba\\
2452419.087 & 4.80 & 130 & 192 & $V$ & Tsukuba\\
2452419.096 & 3.03 & 10 & 698 & -- & Kyoto\\
2452419.114 & 0.10 & 1 & 112 & $R_{\rm c}$ & Ouda\\
2452419.119 & 4.32 & 1 & 5417 & $B$ & Ouda\\
2452419.140 & 3.84 & 3 & 1558 & -- & Okayama\\
2452419.163 & 3.37 & 10 & 740 & -- & Kyoto\\
2452419.183 & 2.50 & 130 & 55 & $B$ & Tsukuba\\
2452419.298 & 2.26 & 10 & 273 & -- & Bronberg\\
2452419.986 & 2.04 & 15 & 371 & -- & Woodridge\\
2452420.083 & 1.61 & 10 & 46 & -- & Tahiti\\
2452420.120 & 0.05 & 1 & 76 & $R_{\rm c}$ & Ouda\\
2452420.123 & 4.06 & 10 & 925 & $B$ & Ouda\\
2452420.137 & 2.05 & 10 & 471 & -- & Kyoto\\
2452420.207 & 2.34 & 10 & 390 & -- & Kyoto\\
2452420.275 & 8.47 & 10 & 1014 & -- & Bronberg\\
2452421.111 & 3.14 & 5 & 1206 & $R_{\rm c}$ & Ouda\\
2452421.272 & 9.12 & 10 & 755 & -- & Bronberg\\
2452422.113 & 4.13 & 5 & 2138 & $R_{\rm c}$ & Ouda \\
2452422.125 & 4.14 & 10 & 868 & -- & Kyoto\\
2452422.125 & 4.14 & 10 & 871 & -- & Kyoto\\
2452424.067 & 3.51 & 10 & 364 & -- & Kyoto\\
2452424.126 & 4.01 & 5 & 1370 & $R_{\rm c}$ & Ouda\\
2452424.523 & 3.58 & 10 & 478 & -- & Bronberg\\
2452426.483 & 2.71 & 15 & 540 & -- & Ceccano\\
2452428.108 & 3.89 & 5 & 1725 & $R_{\rm c}$ & Ouda\\
2452428.131 & 2.93 & 10 & 405 & -- & Kyoto\\
2452429.112 & 3.79 & 5 & 1813 & $R_{\rm c}$ & Ouda\\
2452430.109 & 3.82 & 5 & 1810 & $R_{\rm c}$ & Ouda\\
2452430.561 & 2.50 & 5 & 723 & -- & Okayama\\
2452431.065 & 5.60 & 10 & 1027 & -- & Kyoto\\
2452431.065 & 5.60 & 10 & 1037 & -- & Kyoto\\
2452432.065 & 5.19 & 10 & 809 & -- & Kyoto\\
2452432.112 & 2.57 & 5 & 593 & -- & Okayama\\
2452435.097 & 2.94 & 10 & 372 & -- & Kyoto\\
2452435.111 & 2.35 & 5 & 1157 & $R_{\rm c}$ & Ouda\\
2452437.065 & 2.16 & 30 & 112 & $I_{\rm c}$ & Tsukuba\\
2452437.988 & 4.42 & 30 & 175 & $I_{\rm c}$ & Tsukuba\\
2452437.989 & 3.79 & 80 & 82 & $V$ & Tsukuba\\
2452439.211 & 1.01 & 5 & 142 & $R_{\rm c}$ & Ouda\\
2452439.989 & 3.79 & 80 & 82 & $V$ & Tsukuba\\
2452440.017 & 4.32 & 30 & 176 & $I_{\rm c}$ & Tsukuba\\
2452440.068 & 0.14 & 30 & 14 & $B$ & Tsukuba\\
2452441.070 & 3.86 & 10 & 566 & -- & Kyoto\\
2452442.009 & 3.17 & 130 & 70 & $I_{\rm c}$ & Tsukuba\\
2452442.012 & 2.83 & 30 & 124 & $V$ & Tsukuba\\
2452445.093 & 0.67 & 30 & 58 & $V$ & Tsukuba\\
2452453.076 & 1.49 & 5 & 407 & -- & Okayama\\
2452454.473 & --& 5 & 200 & $V$ & Crimea\\
2452455.403 & --& 5 & 200 & $V$ & Crimea\\
2452456.044 & 1.01 & 10 & 300 & $V$ & Nayoro\\
2452457.996 & 5.72 & 10 & 339 & -- & Kyoto\\
2452458.407 & 0.60 & 200 & 5 & $V$ & Crimea\\
2452459.394 & 0.55 & 200 & 5 & $V$ & Crimea\\
2452461.359 & 1.63 & 200 & 13 & $V$ & Crimea\\
2452462.350 & 1.82 & 200 & 13 & $V$ & Crimea\\
2452462.956 & 3.38 & 20 & 166 & -- & Tahiti\\
2452462.979 & 6.35 & 10 & 515 & -- & Kyoto\\
2452463.078 & 2.90 & 20 & 255 & -- & Wako\\
2452463.114 & 2.11 & 5 & 654 & -- & Okayama\\
2452463.347 & 1.85 & 200 & 13 & $V$ & Crimea\\
2452464.213 & 9.41 & 10 & 993 & -- & Bronberg\\
2452464.342 & 2.42 & 20 & 300 & $V$ & Crimea\\
2452465.048 & 0.22 & 5 & 88 & $V$ & Nayoro\\
2452466.297 & 7.56 & 10 & 1057 & -- & Bronberg\\
2452466.649 & 0.30 &  5 & 100 & -- & Seikei\\
2452466.978 & 6.41 & 10 & 655 & -- & Kyoto\\
2452467.004 & 2.54 & 15 & 462 & -- & Wako\\
2452467.909 & 6.67 & 15 & 1199 & -- & Woodridge\\
2452467.959 & 4.41 & 10 & 467 & -- & Kyoto\\
2452468.077 & 2.90 & 15 & 264 & -- & Wako\\
2452468.423 & 1.82 & 7 & 722 & -- & Ceccano\\
2452468.895 & 3.96 & 20 & 360 & -- & Tahiti\\
2452469.284 & 7.68 & 10 & 1086 & -- & Bronberg\\
2452471.964 & 4.94 & 5 & 738 & $R_{\rm c}$ & Ouda\\
2452471.982 & 4.01 & 10 & 258 & -- & Kyoto\\
2452472.011 & 2.38 & 15 & 274 & -- & Wako\\
2452475.972 & 0.80 & 10 & 117 & -- & Kyoto\\
2452476.061 & 3.43 & 5 & 513 & -- & Wako\\
2452476.971 & 5.67 & 10 & 787 & -- & Kyoto\\
2452479.048 & 1.49 & 10 & 453 & -- & Kyoto\\
2452480.984 & 5.20 & 10 & 683 & -- & Kyoto\\
2452481.985 & 4.94 & 10 & 713 & -- & Kyoto\\
2452491.954 & 5.09 & 10 & 1120 & -- & Kyoto\\
2452492.959 & 4.68 & 10 & 807 & -- & Kyoto\\
2452493.969 & 4.33 & 10 & 872 & -- & Kyoto\\
2452496.124 & 0.63 & 10 & 137 & -- & Kyoto\\
2452497.949 & 4.47 & 10 & 790 & -- & Kyoto\\
2452498.957 & 0.74 & 10 & 155 & -- & Kyoto\\
2452499.952 & 4.46 & 10 & 903 & -- & Kyoto\\
2452502.968 & 4.18 & 10 & 820 & -- & Kyoto\\
2452508.944 & 3.65 & 10 & 365 & -- & Kyoto\\
2452511.954 & 3.67 & 10 & 684 & -- & Kyoto\\
2452512.935 & 2.30 & 10 & 405 & -- & Kyoto\\
2452515.047 & 1.20 & 10 & 258 & -- & Kyoto\\
2452515.952 & 3.30 & 10 & 701 & -- & Kyoto\\
2452516.936 & 3.62 & 10 & 662 & -- & Kyoto\\
2452518.937 & 3.68 & 10 & 817 & -- & Kyoto\\
2452519.934 & 3.64 & 10 & 571 & -- & Kyoto\\
2452522.928 & 3.43 & 10 & 704 & -- & Kyoto\\
2452525.929 & 3.38 & 10 & 716 & -- & Kyoto\\
2452526.914 & 3.78 & 10 & 798 & -- & Kyoto\\
2452535.915 & 2.73 & 10 & 554 & -- & Kyoto\\
2452536.912 & 2.96 & 10 & 493 & -- & Kyoto\\
2452537.926 & 2.07 & 10 & 141 & -- & Kyoto\\
2452540.907 & 2.80 & 10 & 615 & -- & Kyoto\\
2452541.944 & 2.93 & 10 & 556 & -- & Kyoto\\
2452557.880 & 2.33 & 10 & 513 & -- & Kyoto\\
2452558.899 & 1.74 & 10 & 362 & -- & Kyoto\\
2452559.922 & 0.98 & 10 & 203 & -- & Kyoto\\
2452560.883 & 2.08 & 10 & 454 & -- & Kyoto\\
2452561.873 & 2.26 & 10 & 498 & -- & Kyoto\\
2452563.882 & 2.02 & 10 & 446 & -- & Kyoto\\
2452564.876 & 1.67 & 10 & 362 & -- & Kyoto\\
2452568.882 & 1.29 & 10 & 234 & -- & Kyoto\\
2452569.871 & 1.57 & 10 & 299 & -- & Kyoto\\
\hline
\end{longtable}

\section{Results}
\subsection{Outburst in 2002 May}

\begin{figure*}
  \begin{center}
    \FigureFile(170mm,170mm){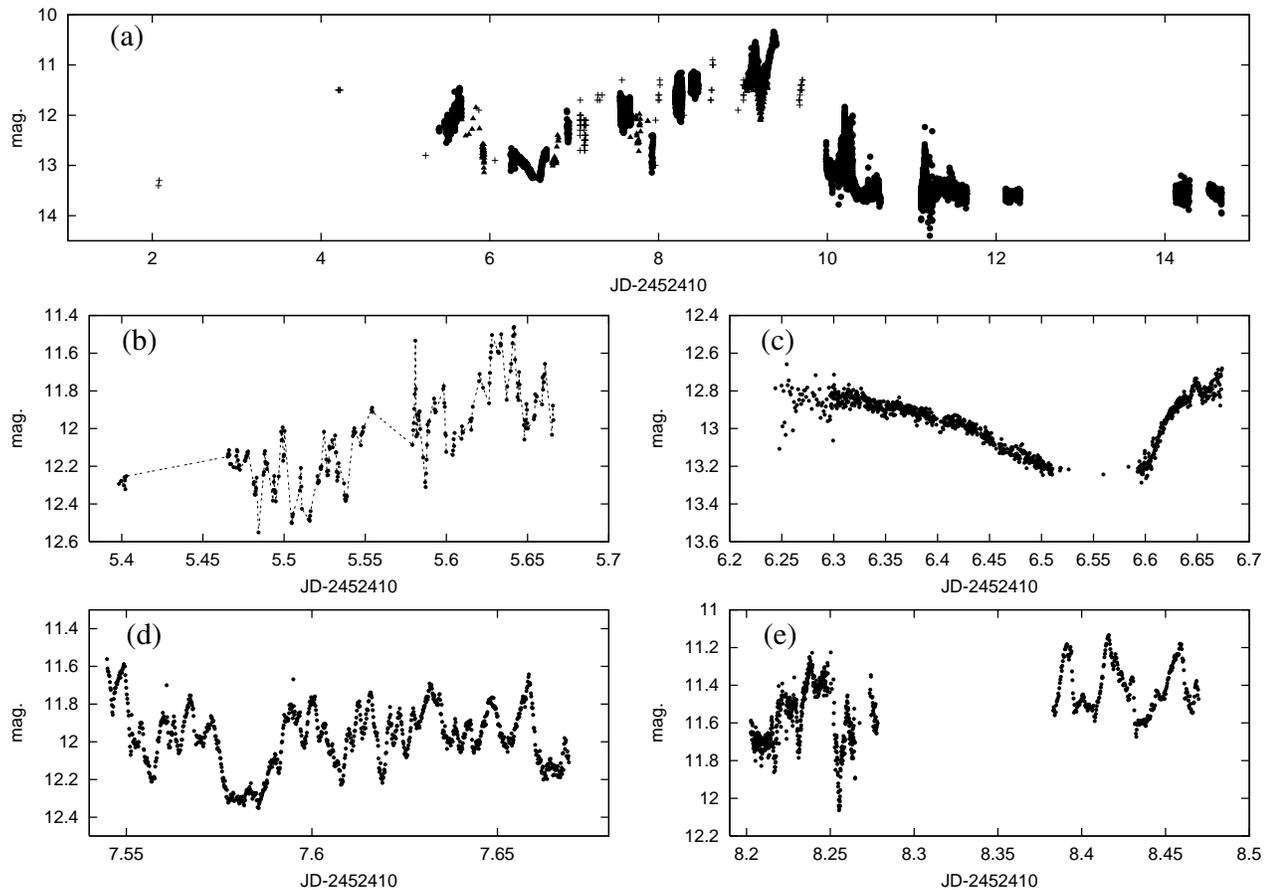}
  \end{center}
  \caption{Light curves during the May outburst.  The abscissa and
 ordinate denote the time in JD and the $R_{\rm c}$ (and $V$ in the
 panel (a)) magnitude, respectively.  Panel (a): The whole light curve of
 the outburst.  The filled circles and triangles are $R_{\rm c}$- and
 $V$-magnitudes from our CCD observations, respectively.  The crosses
 show visual observations reported to VSNET.  Panel (b), (c), (d), and (e):
 Short-term variations during the outburst.  Typical errors
 of each observation is 0.02 mag in the panel (b), 0.03 mag in the panel
 (c), 0.02 mag in the panel (d), 0.04 mag in the early observation in the
 panel (e), and 0.02 mag in the late observation in the panel (e).  In the
 panel (b), we also show observations with a dotted line to show
 short-term variations more clearly.}
\label{fig:out}
\end{figure*}

The outburst in May is characterized by a short total duration of about
6 days and multiple peaks as reported in \citet{uem02v4641letter}.  The
whole light curve of the outburst is shown in the panel (a) of figure
\ref{fig:out}.  In this figure, we also show examples of rapid
variations detected during the outburst in the panel (b), (c), (d), and
(e).  The abscissa is the time in JD and the ordinate is the $R_{\rm c}$
magnitude.

One day after the first visual detection of the outburst (11.5 mag on JD
2452414), the object was detected at a fainter level of 12.8 mag.
A subsequent rapid brightening on JD 2452415 is shown in the panel (b).
The light curve shows short-term fluctuations with a time scale of 100 s
and with large amplitudes of 0.1--0.5 mag.  After this epoch, both
visual and CCD observations recorded a temporal fading of about 1 mag.
The tail of the fading and the rapid recovering was detected by our
time-series observation on JD 2452416, as shown in the panel (c).  It is
notable that the strong short-term fluctuations in the panel (b) almost
disappeared during this phase.  A regrowth of short-term variations can
be seen at the end of the light curve in the panel (c).  On JD 2452417,
the object seems to have remained at a bright state at about 12 mag.
The light curve in the panel (d) shows more rapid short-term variations
of a time scale of $\sim 30\;{\rm s}$.  It is also characterized by a
broad dip around JD 2452417.58, during which the short-term variation
was relatively weak.  The object again experienced a temporal fading,
and then quickly recovered.  The light curve in early JD 2452418 (panel
(e)) was dominated by several dips rather than the fluctuations seen in
the panel (b) and (d).  The duration of the dips were only about 500 s,
and the deepest dip was about 0.7 mag.  These dips were, however, not
observed 0.1 day after this epoch.  As can be seen in the panel (e), the
light curve in the middle JD 2452418 shows rather smooth variations of
$\sim 1000\;{\rm s}$.

\begin{figure*}
  \begin{center}
    \FigureFile(170mm,170mm){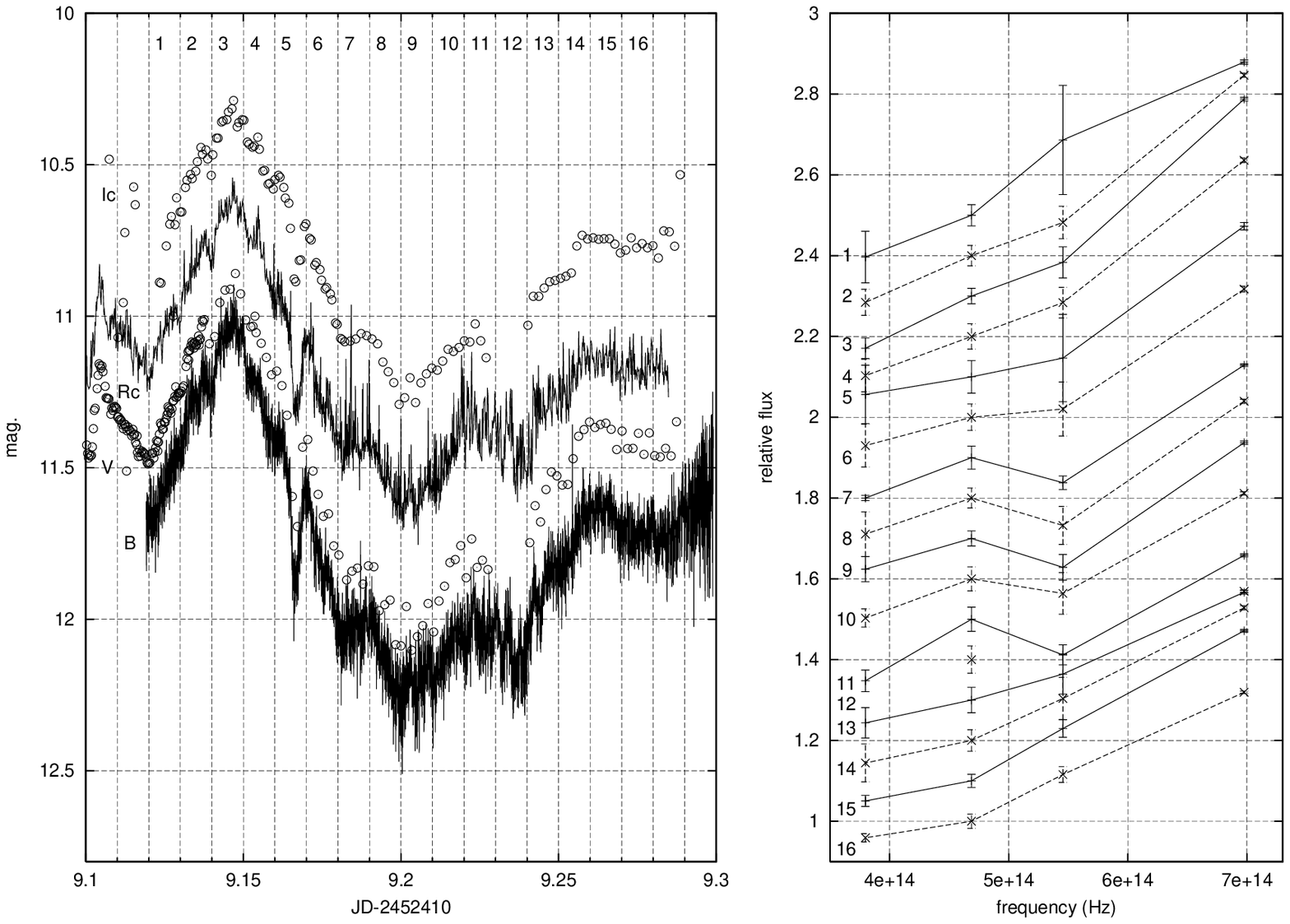}
  \end{center}
  \caption{$B$, $V$, $R_{\rm c}$, and $I_{\rm c}$ simultaneous
 observations around the maximum of the outburst (JD 2452419).  Left panel:
 Multi-color light curves.  The abscissa and ordinate denote the time in
 JD and the magnitude, respectively.  We show the light curves of
 $I_{\rm c}$ (open circles), $R_{\rm c}$ (line), $V$ (open circles), and
 $B$ (line) from top to bottom.  Right panel: Time evolution of the
 de-reddened spectral energy distribution.  The abscissa and ordinate
 denote the frequency in Hz and the relative flux in an arbitrary unit
 which is normalized at the $R_{\rm c}$-band flux, respectively.  The
 left-side number with each spectrum means the time bin of
 $\Delta t=0.01\;{\rm d}$ shown in the left panel.  The error is the
 standard error in each bin.  In the bin No. 12, only $B$ and 
 $R_{\rm c}$-band observations are available.  The $R_{\rm c}$-band 
 observation is shown as an isolated point in the right panel, and the
 $B$-band observation is roughly coincident with that of the bin
 No. 13.}
\label{fig:0524sed}
\end{figure*}

The object reached the maximum of the outburst on JD 2452419.  We
succeeded in obtaining multi-color light curves just before the maximum.
The left panel of figure \ref{fig:0524sed} shows the resulting light
curves of $B$, $V$, $R_{\rm c}$, and $I_{\rm c}$ bands.  We divided the
light curve into 16 bins of $\Delta t=0.01\;{\rm d}$, as labeled in the
figure, and calculated de-reddened, mean flux of each bin.  The obtained
spectral energy distributions (SEDs) are normalized with the $R_{\rm
c}$-flux and shown in the right panel of figure \ref{fig:0524sed}.  The
de-reddening procedure was performed with $E(B-V)=0.32$ reported in
\citet{oro01v4641sgr} and the relationship on $E(V-R_{\rm c})$ and
$E(R_{\rm c}-I_{\rm c})$ reported in \citet{tay86color}.

The object experienced a large hump in all bands during the bin No. 1--6
in figure \ref{fig:0524sed}.  The peak was in the bin No. 3, when the
object became bluer, as can be seen in the right panel of figure
\ref{fig:0524sed}.  After the hump, short-term fluctuations with a time
scale of 100 s became stronger (the bin No. 9, 11, and 12), as can be
seen in $B$- and $R_{\rm c}$-band light curves.  The $V$- and $I_{\rm
c}$-band light curves are relatively sparse, however the rapid
variations can be confirmed even in these bands.  The SED during this
phase is characterized by flat spectra in the $V$--$I_{\rm c}$ region
with the probable excess of the $R_{\rm c}$-flux.  The flat spectral
component again weakened since the bin No. 13, when another brightening
started.  As can be seen from figure \ref{fig:out} and
\ref{fig:0524sed}, the properties of rapid variations were drastically
changed within a short period, which was typically a few hours.

\begin{figure*}
  \begin{center}
    \FigureFile(170mm,170mm){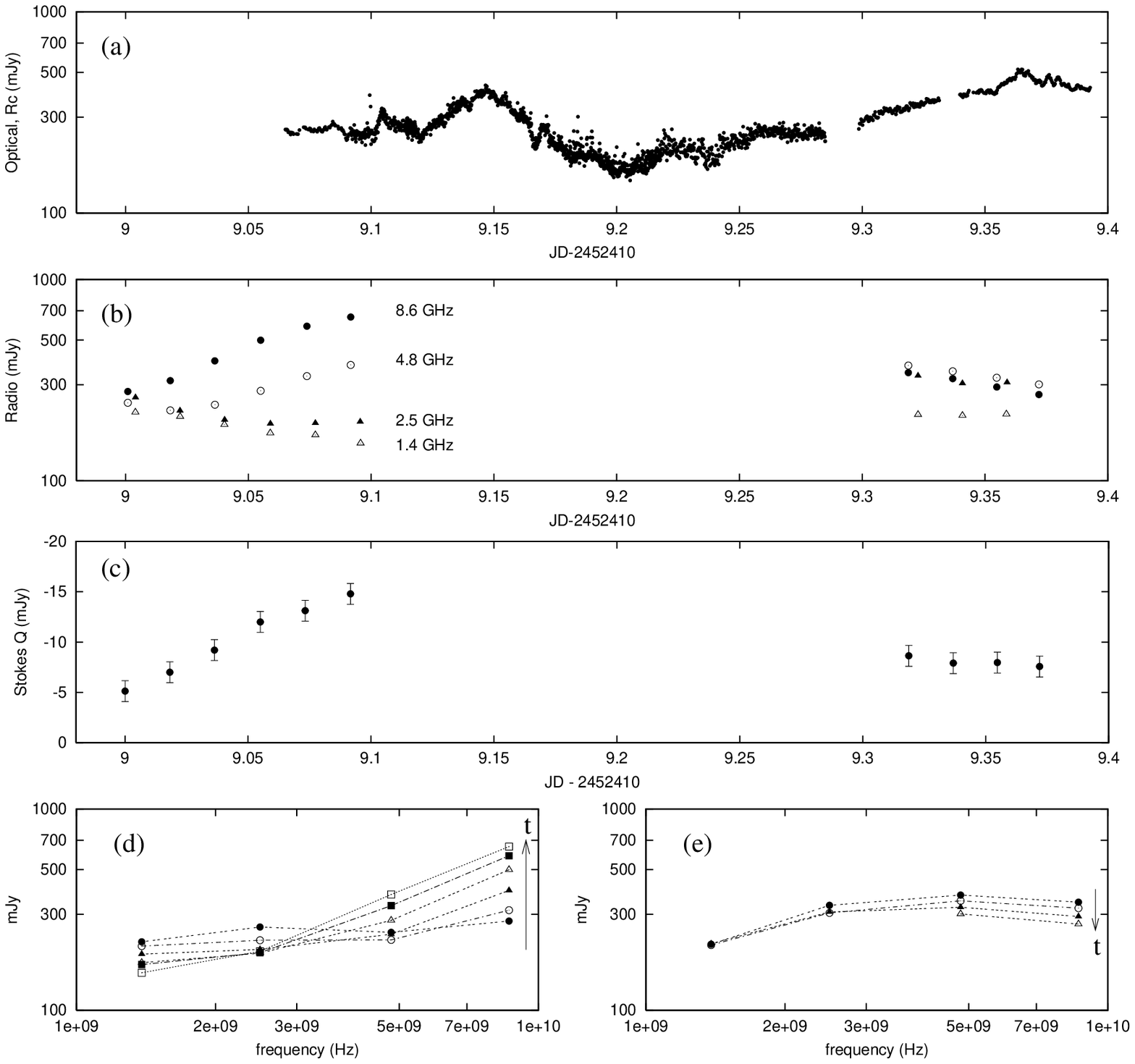}
  \end{center}
  \caption{Optical and radio observations around the maximum of the
 outburst (JD 2452419).  Panel (a): Optical $R_{\rm c}$-band light curve.
 The abscissa and ordinate are the time in JD and the flux density in mJy,
 respectively.  Extinction corrections to the optical data were
 performed.  Panel (b): Radio light curves.  The abscissa and ordinate
 are same as the panel (a).  The filled circles, the open circles, the
 filled triangles, and the open triangles denote the radio observations
 at 8.64, 4.8, 2.496, and 1.384 GHz, respectively.  The standard errors
 of each point are smaller than the symbol size in the figure.  Panel
 (c): Time evolution of the Stokes Q at 8.64 GHz.  Panel (d) and (e): Time
 evolution of spectral energy distributions in the radio range.  The
 abscissa and ordinate denote the frequency in Hz and the flux density in mJy,
 respectively.  Observations in different epoch are shown with distinct
 symbols.} 
\label{fig:0524radio}
\end{figure*}

Figure \ref{fig:0524radio} shows the result of radio observations on JD
2452419, a part of which was simultaneously performed with the optical
observation.  The panel (a) and (b) describe the optical and radio
variations, respectively.  The light curve in the panel (a) is the
optical one corrected with the extinction of $A_{R_{\rm c}}=0.86$ which
was estimated from $E(B-V)=0.32$.  In the panel (b), the filled circles,
open circles, filled triangles, and open triangles denote the radio flux
of 8.64, 4.80, 2.50, and 1.38 GHz, respectively.  No polarization was
detected above the $3\,$mJy level, except at 8.64\,GHz on JD~2452419,
where Stokes $Q$ amplitude was observed to peak, with $Q=-16\,mJy$ as
can be seen in the panel (c), and subsequently decay as the total
intensity decayed.  The SEDs in the radio range are shown in the panel
(d) for the early six sets and (e) for the late four sets of
observations.  We calculated the spectral index $\alpha$ defined with
$f_{\nu}\propto \nu^\alpha$, using all points of each epoch and three
high frequency points (8.64, 4.80, and 2.50 Hz).  The results are
presented in table \ref{tab:0524radio}.  We calculated two types of
$\alpha$ because there is a possible break at 2--3 GHz in some SEDs.

As can be seen in figure \ref{fig:0524radio} and table
\ref{tab:0524radio}, the radio SEDs were first rapidly changing from
almost flat spectra to highly inverted spectra.  The spectral index
reached at a surprisingly high level of about 0.8--1.0 around JD
2452419.1.  In the late four spectra, the high-frequency flux gradually
faded and the low-frequency flux almost remained constant.  The spectra
again became flat during this period.  There is no clear correlation
between the optical and the radio light curves within our available
data.  

\begin{table*}
\caption{Radio spectral index}
\label{tab:0524radio}
\begin{center}
\begin{tabular}{ccc}
\hline \hline
JD$-$2452420 & $\alpha$ (from all data) & $\alpha$ (from 3 high frequencies)\\
\hline
9.003 & 0.11$\pm$ 0.05 & 0.05$\pm$ 0.09\\
9.020 & 0.20$\pm$ 0.09 & 0.27$\pm$ 0.16\\
9.038 & 0.38$\pm$ 0.12 & 0.54$\pm$ 0.17\\
9.057 & 0.58$\pm$ 0.12 & 0.77$\pm$ 0.12\\
9.076 & 0.70$\pm$ 0.12 & 0.89$\pm$ 0.04\\
9.094 & 0.81$\pm$ 0.09 & 0.97$\pm$ 0.02\\
\hline
9.321 & 0.25$\pm$ 0.14 & 0.03$\pm$ 0.09\\
9.339 & 0.23$\pm$ 0.12 & 0.05$\pm$ 0.10\\
9.357 & 0.16$\pm$ 0.12 & -0.04$\pm$ 0.08\\
9.372 & -0.20 & -0.20\\
\hline
9.928 & -0.25$\pm$ 0.09 & -0.34$\pm$ 0.16\\
10.934 & -0.66$\pm$ 0.09 & -0.66$\pm$ 0.09\\
\hline
\end{tabular}
\end{center}
\end{table*}

\begin{figure}
  \begin{center}
    \FigureFile(85mm,85mm){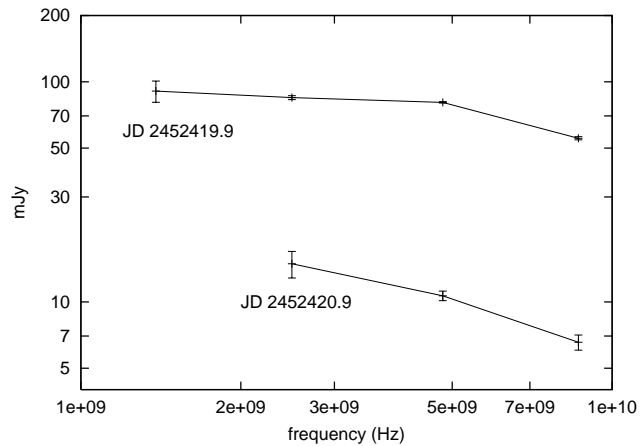}
  \end{center}
  \caption{Spectral energy distributions in the radio range just after
 the outburst.  The abscissa and ordinate denote the frequency in Hz and
 the flux density in mJy, respectively.  The upper and lower points are
 observations on JD 2452419.9 and JD 2452420.9, respectively.  The error
 in the figure is the standard error.}
\label{fig:052526radio}
\end{figure}

The outburst was terminated by a sudden fading in the end of JD
2452419.  The object then entered a post-outburst active phase.  Both
the radio flux and the spectral index rapidly decreased with time as
shown in table \ref{tab:0524radio} and figure \ref{fig:052526radio}.

\subsection{Post-outburst active phase}

\subsubsection{Correlation between optical and X-ray emission just after
   the outburst}

As reported in \citet{uem02v4641letter}, our optical observation
detected repeated flaring on JD 2452420, just after the outburst.  This
active phase has notable characteristics compared with known optical
activities observed in other black hole binaries.  First, the flares
have large amplitudes of $\gtrsim 1$ mag (peak magnitudes $\sim 11.9$
mag) with short time scales.  On their time scale, the $e$-folding time
of rising or fading branches is calculated to be typically $\sim 10\;{\rm
min}$, which is quite shorter than the time scale at the outermost
region of the accretion disk.  Second, the duration of this active phase
was only about 3.5 hr, and it was suddenly terminated by a subsequent
calm state.  In the X-ray range, such rapid state transitions are well
known in GRS 1915+105 (\cite{gre96grs1915}; \cite{taa97grs1915};
\cite{yad99grs1915}).  V4641 Sgr is a unique source in the point that
such rapid state transitions were observed in the optical range.

\begin{figure*}
  \begin{center}
    \FigureFile(170mm,170mm){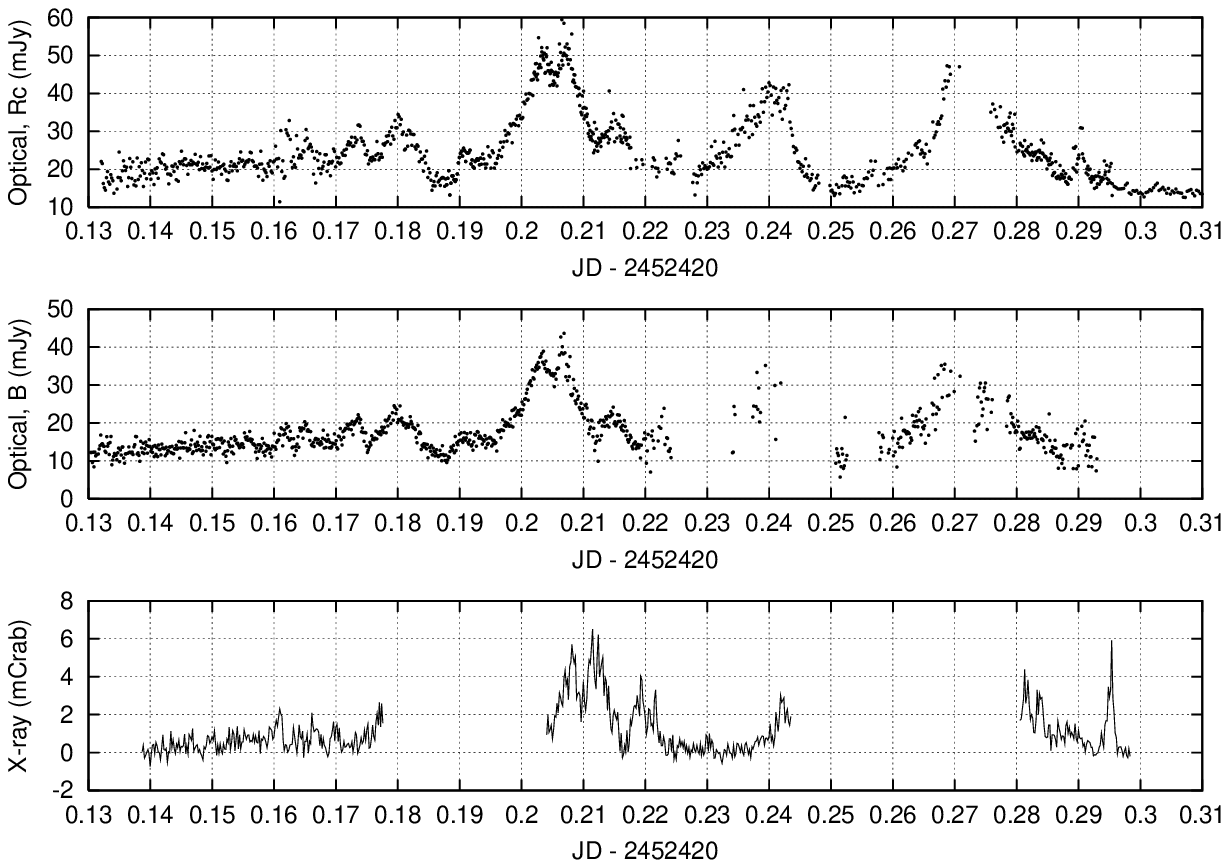}
  \end{center}
  \caption{Optical and X-ray light curves just after the outburst (JD
 2452420).  The abscissa and ordinate denote the time in JD and the flux
 density in mJy in the optical light curves and in mCrab in the X-ray light
 curve.  Top panel: Optical $R_{\rm c}$-band light curve.  Middle panel:
 Optical $B$-band light curve.  Interstellar extinction corrections were
 performed to these optical data.  The typical error is 2.0 mJy in the
 $R_{\rm c}$-band data and 1.5 mJy in the $B$-band data.  Bottom panel:
 X-ray light curve.  The typical error is 0.3 mCrab.  }
\label{fig:0525x}
\end{figure*}

During the active phase on JD 2452420, we succeeded in obtaining
simultaneous light curves of optical ($R_{\rm c}$ and $B$) and
X-ray emission.  The light curves are shown in figure \ref{fig:0525x}.
The top, middle, and bottom panels are the light curve of $R_{\rm
c}$-band, $B$-band, and X-ray flux, respectively.  In these light
curves, the abscissa denotes the time in geocentric JD.  The ordinate
denotes the flux density in mJy for the optical data and in mCrab for the X-ray
data.

Our observations demonstrate a clear correlation between the optical and
the X-ray light curves.  The first giant flare around the time of JD
2452420.19--2452420.23 in figure \ref{fig:0525x} is connected with a
large double-peaked optical flare. Subsequent small optical flares
during the same interval preceded several X-ray flares with profiles
that resembled those observed at optical wavelengths.  The time scale of
the X-ray flare is shorter than that of the optical one.  We calculated
$e$-folding times of rising branches of this flare to be $12\pm 1\;{\rm
min}$ in the optical range and $3\pm 1\;{\rm min}$ in the X-ray range.
Compared with the optical light curve, rapid fluctuations superimposed
on the giant flare are more prominent in the X-ray light curve.  In the
optical range, the object became bluer and reached $B-R_{\rm c}\sim 0.6$
at the peak.  We obtain the de-reddened color of $B-R_{\rm c}\sim 0.1$
with the reported reddening of $E(B-V)=0.32$ (\cite{oro01v4641sgr}).

As well as this giant flare, we can see correlated X-ray--optical
variations even in small flares around JD 2452420.15--2452420.18, in a
rising to the second giant flare around JD 2452420.23--2452420.25, and
in a short flare around JD 2452420.29--2452420.30.  In Table
\ref{tab:0525cor}, we present delay times calculated by
cross-correlations.  As shown in this table, the X-ray delay is
typically $\sim$7 min, while the delay times have possible dispersions.
The detection of the X-ray delay means that the optical emission was
definitely not a reprocess of the X-ray emission.  The table also shows
marginal detection of $B$-band delays of about 20 s.

\begin{table*}
\caption{X-ray and $B$-band flux delays against $R_{\rm c}$-band flux calculated with cross-correlations.}
\label{tab:0525cor}
\begin{center}
\begin{tabular}{cccc}
\hline \hline
JD$-$2452420 & X-ray$-$Optical($R_{\rm c}$) (min)& X-ray$-$Optical($B$) (min)&
 Optical($B$)$-$Optical($R_{\rm c}$) (min)\\
\hline
0.15--0.18 & $5.67\pm 0.98$ & $6.40\pm 0.70$ & $0.34\pm 0.21$ \\
0.19--0.23 & $7.25\pm 0.42$ & $7.01\pm 0.37$ & $0.29\pm 0.19$ \\
0.28--0.30 & ---            & $6.85\pm 0.72$ & --- \\
\hline
\end{tabular}
\end{center}
\end{table*}

\begin{figure*}
  \begin{center}
    \FigureFile(170mm,170mm){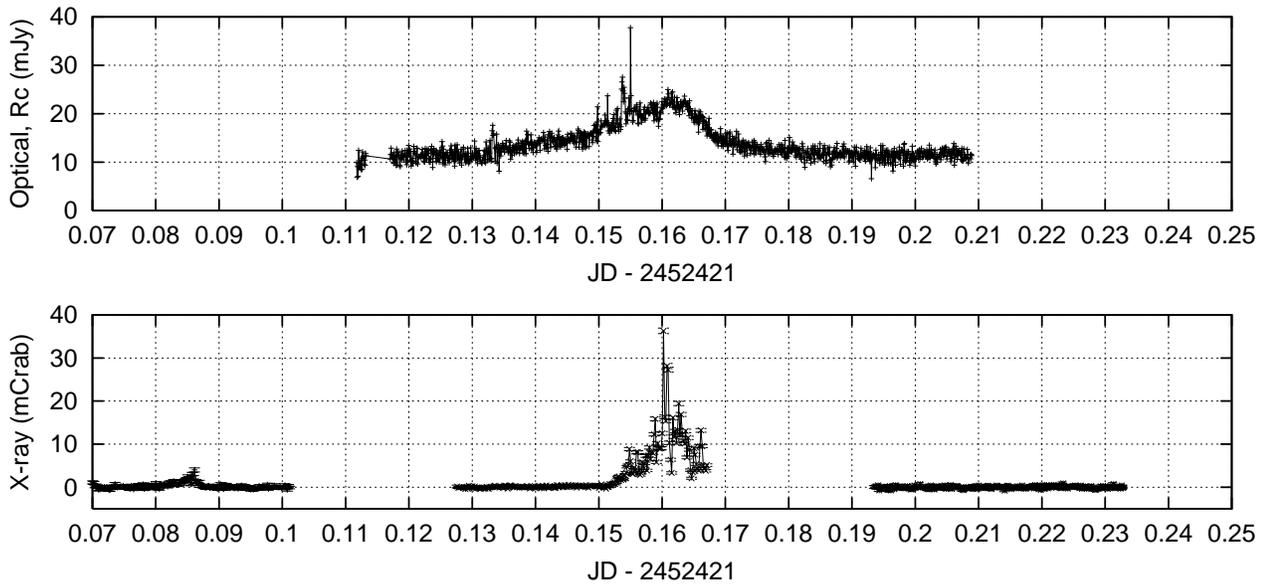}
  \end{center}
  \caption{Optical and X-ray light curves just after the outburst (JD
 2452421).  The abscissa and ordinate are same as figure
 \ref{fig:0525x}.  Upper panel: Optical $R_{\rm c}$-band light curve.
 Interstellar extinction corrections were performed.  The typical error
 is 1.1 mJy in the optical data.  Bottom panel: X-ray light curve.  The
 typical error is 0.2 mCrab during the low level and 0.4 mCrab during
 the flare.}
\label{fig:0526x}
\end{figure*}

Simultaneous X-ray and optical observations were also performed on
JD~2452421.  The resulting light curves are shown in figure
\ref{fig:0526x}.  The abscissa and ordinate are same as figure
\ref{fig:0525x}.  The object was still active on JD 2452421.  We can see
a smaller optical flare in the upper panel of figure \ref{fig:0526x}.
Compared with flares on JD 2452420, the flare on JD 2452421 has a longer
duration of $\sim 1\;{\rm hr}$ and a larger $e$-folding time of rising
branch of $62\pm 2$ min.

Between JD 2452421.15--JD 2452421.16 in figure \ref{fig:0526x}, our
observation detected possible strong fluctuations superimposed on the
more gradual trend of optical flare.  We carefully checked images around
these short-term variations and confirmed no change of transparency.  At
least one brightening which occurred at JD 2452421.154 was recorded in
several images, which supports that they are real events.

In the X-ray range, a corresponding flare was also detected.  It started
rising at JD 2452421.152, which is about 24 min after the onset of the
optical flare.  An X-ray delay was hence observed also on JD 2452421 as
well as the flares on JD 2452420, and furthermore, it became longer.
This X-ray flare is characterized by strong fluctuations on the more
gradual flare component.  As can be seen in figure \ref{fig:0526x}, the
X-ray flux first increased rather smoothly, and then, the fluctuations
became stronger with time.  The fluctuation was strongest around JD
2452421.16 when the light curve was dominated by striking rapid
brightenings.  Some of them recorded over 20 mCrab.  The X-ray peak
luminosity on JD 2452421 is much higher than that on JD 2452420.  The
observed peak time of the X-ray flux roughly coincides with the optical
peak.  This is a notable difference compared with the flares on JD
2452420, in which both risings and peaks of the optical flares preceded
those of the X-ray flux.

\subsection{Active phase after the outburst}

\begin{figure*}
  \begin{center}
    \FigureFile(170mm,170mm){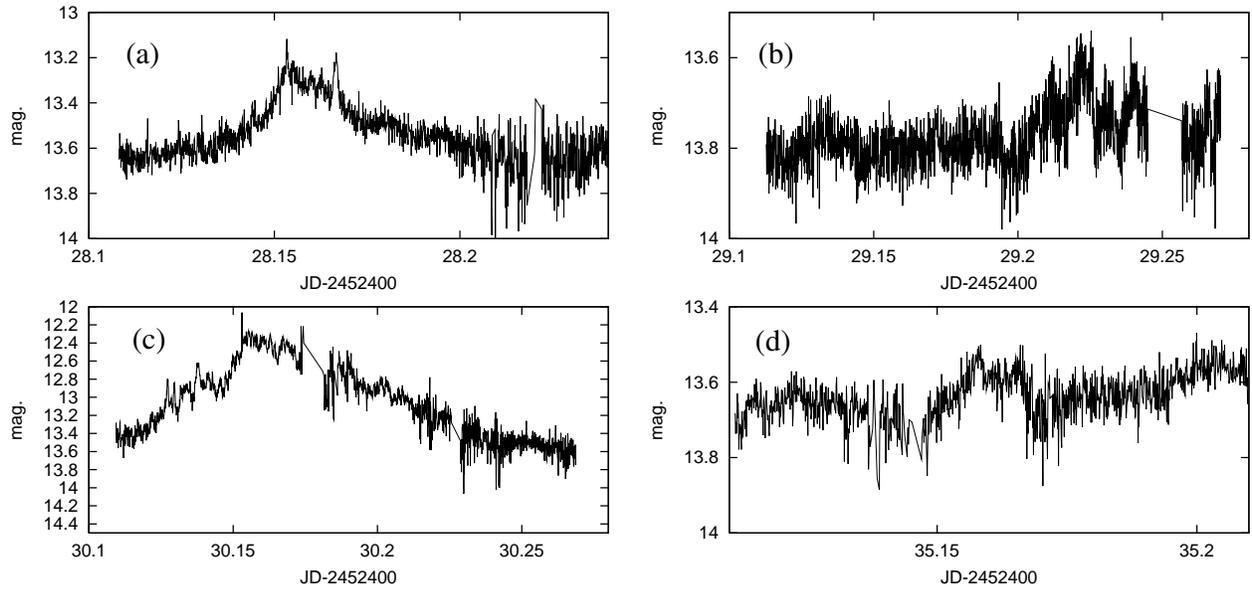}
  \end{center}
  \caption{Light curves showing rapid variations after the outburst.
 The abscissa and ordinate denote the time in JD and the $R_{\rm c}$
 magnitude, respectively.  The typical error is 0.04 mag in the panel (a),
 (b), and (d), and 0.07 mag in the panel (c).  The power spectra of these
 variations are shown in figure \ref{fig:powers}. }
\label{fig:active}
\end{figure*}

Our optical observations detected clear brightenings and fluctuations
even after JD 2452421.  They are shown in figure \ref{fig:active}.  A
large flare with the peak magnitude of 12.2 mag was detected on JD
2452430, as shown in the panel (c) of figure \ref{fig:active}.  While
its peak magnitude is comparable to those of flares observed on JD
2452420 and 2452421, the duration is much longer, at least 0.12 d.
Other three panels (a, b, and d) show rapid variations with amplitudes
smaller than 1 mag.  They are rather minor, however evidently indicate
that the object had remained active at least for two weeks after the
outburst. 

\begin{figure}
  \begin{center}
    \FigureFile(85mm,85mm){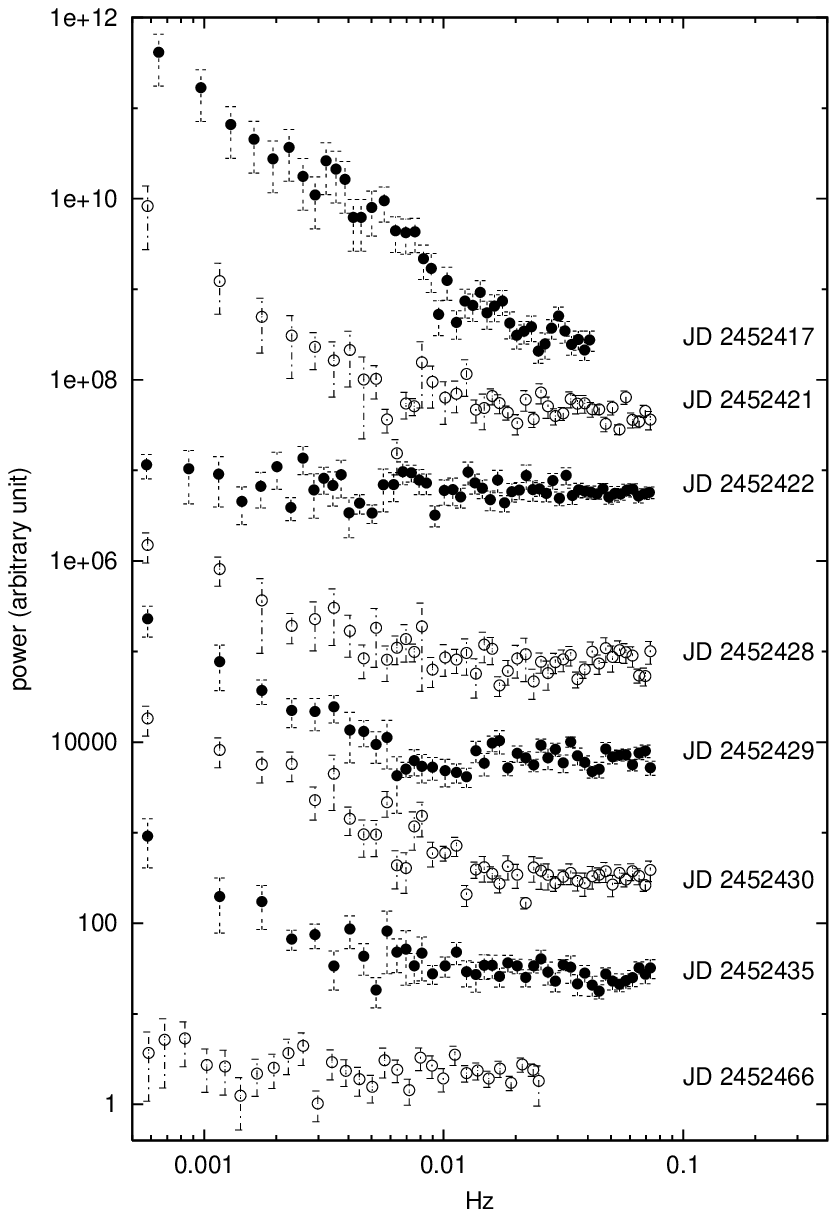}
  \end{center}
  \caption{Power spectra of optical short-term variations.  The abscissa
 and ordinate are the frequency in Hz and the power in an arbitrary unit.
 The top points with the filled circles are a power spectrum during the
 outburst (JD 2452417).  The other spectra are those after the
 outburst.  The power spectra on JD 2452428--2452435 correspond to the
 short-term variations shown in figure \ref{fig:active}.}
\label{fig:powers}
\end{figure}

Figure \ref{fig:powers} shows examples of power spectra of variations
during and after the outburst.  The abscissa and ordinate are the
frequency in Hz and the power in an arbitrary unit.  As can be seen in
these power spectra, short-term variations of $\sim 100\;{\rm s}$ still
appeared after the outburst as well as $\sim 1000$ s variations.  On JD
2452422, the object was calm throughout our observation for 0.17 d, as
can be seen in the power spectrum in figure \ref{fig:powers}.  It is,
however, possible that a number of similar variations were overlooked
due to their short durations.  Within our available data, no other rapid
variation with amplitudes over 0.2 mag was detected until JD 2452462,
when the object again became active as reported in the next section.

\subsubsection{Optical flash on July 7}

\begin{figure*}
  \begin{center}
    \FigureFile(170mm,170mm){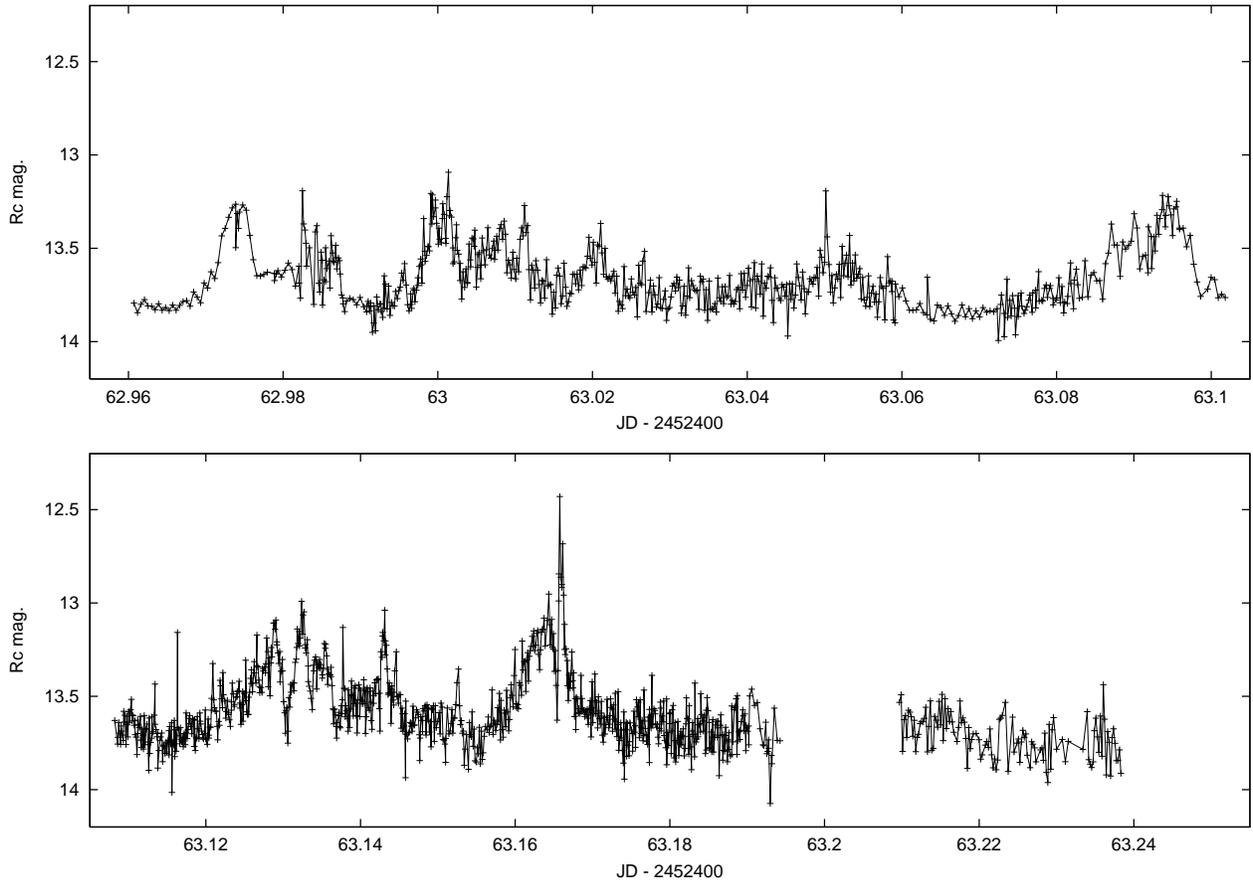}
  \end{center}
  \caption{Light curves of V4641 Sgr on 2002 July 7. The abscissa and
 ordinate denote the date and $R_{\rm c}$ magnitude, respectively.  The
 error of each point is typically 0.05 mag.}
\label{fig:flash}
\end{figure*}

On JD 2452462--2452463, the object exhibited sporadic, large-amplitude
brightenings, as shown in figure \ref{fig:flash}, while it almost
remained at the quiescent level except for the period of the
brightenings.  The activity became stronger with time; we can see
brightenings with amplitudes of $\sim 0.5$ mag and durations of 100--500
s in the upper panel of figure \ref{fig:flash}, and then, remarkable
rapid brightenings with a typical duration of 10--100 s in the lower
panel.  These ``optical flashes'' were superimposed on more gradual
brightenings of about 0.5 mag.  We cannot detect any periodicity in
them.

\begin{figure}
  \begin{center}
    \FigureFile(85mm,85mm){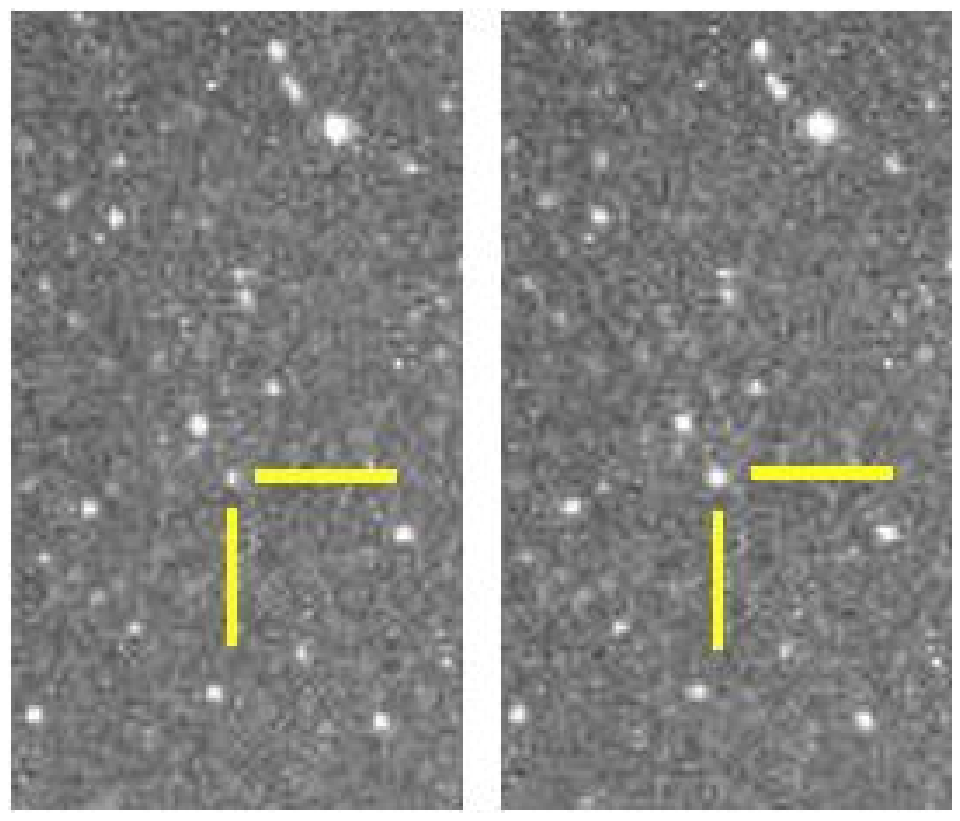}
  \end{center}
  \caption{CCD images of V4641 Sgr just before (left) and during (right)
 the giant optical flash.  The images were observed at JD 2452463.16542
 (left) and 2452463.16577.  The time separation from the left to right
 images is 30 s. Each frame is $2^\prime .5\times 4^\prime .3$; north is
 up and east to the left.}
\label{fig:image}
\end{figure}

The most prominent flash occurred around JD 2452463.165--2452463.167.
During this flash, the object brightened to the maximum of $12.4\;{\rm
mag}$ within 30 s, and then, returned to the pre-flash level within 90
s.  Although the duration of the flash was very short, our observations
successfully detected several points just during both ascending and
descending branches, which provide unambiguous evidence for these
variations.  Figure \ref{fig:image} shows CCD images taken just before
and during this flash.  We can easily confirm the brightening at a high
confidence level, comparing V4641 Sgr (the marked object) with neighbor
stars.  The optical flashes were detected only in the period shown in
figure \ref{fig:flash} and no similar variation was confirmed before and
after this period.

As well as the optical flashes, several dips just before rapid
brightenings are remarkable.  Clear dips occurred at JD 2452463.130 and 
JD 2452463.165 just before the most prominent flash.  We can see some
another possible dips during the brightening periods, whereas no
dip-like feature was detected when the object was at the quiescent
level. 

Radio observations show that no source was found on June 5 at the
position of V4641 Sgr.  A radio source again appeared on July 13, and
then, remained active on July 16, 24, 27, 28, and 29
(\cite{rup02v4641iauc}).\footnote{$\langle$http://vsnet.kusastro.kyoto-u.ac.jp/vsnet/Mail/vsnet-campaign-v4641sgr/msg00102.html,
msg00104.html$\rangle$}  No X-ray observation is available around the
optical flashes.\footnote{$\langle$http://xte.mit.edu/$\rangle$}  After
JD 2452463, no short-term variation with amplitudes over 0.2 mag was
detected within our available data.   It was reported that the X-ray
luminosity was lower than $10^{32}\;{\rm erg\,s^{-1}}$ on August 5,
which indicates that the object had returned to the quiescent state
(\cite{tom02v4641ATEL}).

\subsection{Quiescent state}

While several rapid brightenings were detected by our observation even
after the May outburst, the object quickly returned to the quiescent
level in the optical range just after the outburst.  We performed period
analysis using observations in 2002 combined with the past data reported
in \citet{uem02v4641}.  The data contains 337-night observations from JD
2437109 to 2452569.  We excluded light curves during the outburst or
clearly showing rapid variations.  Heliocentric corrections to the
observed times were applied before the period analysis.  Our period
analysis with the phase dispersion minimization method (PDM; \cite{PDM})
yields the best period of $P=2.817280\pm 0.000015\;{\rm d}$, which is in
agreement with the period reported in \citet{uem02v4641} and
\citet{oro01v4641sgr}.

\begin{figure}
  \begin{center}
    \FigureFile(85mm,85mm){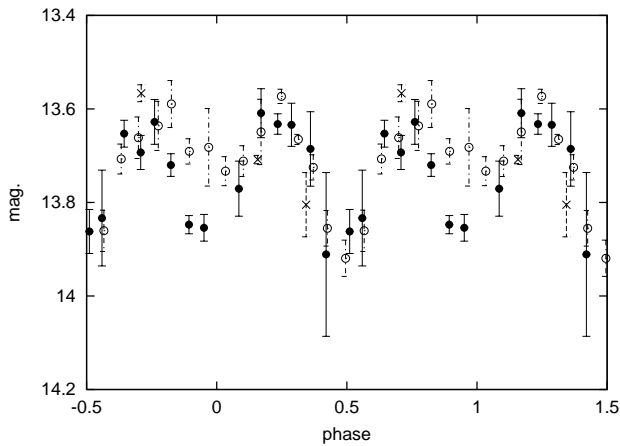}
  \end{center}
  \caption{Ellipsoidal modulations observed after the outburst in 2002
 (filled circles) and 1999--2001 (open circles; \cite{uem02v4641}).  The
 abscissa denotes the phase calculated with the period of 2.81728 d and
 the epoch of HJD 2447708.89515 (\cite{oro01v4641sgr}).  The ordinate
 denotes $R_{\rm c}$-magnitude.  The three crosses are
 observations in 2002, just before the outburst (JD 2452397, 2452406,
 and 2452407).}
\label{fig:quies}
\end{figure}

Figure \ref{fig:quies} shows ellipsoidal modulations in 2002
June--October (JD 2452492--JD 2452569; filled circles) and in 1999--2001
(\cite{uem02v4641}; open circles).  As can be seen in the figure, the
object exhibited ordinary ellipsoidal modulations after the outburst, as
observed in 1999--2001.  It is interesting to note that the object is
possibly fainter around the secondary minimum after the May outburst,
whereas the two maxima and the primary minimum are in agreement with
those in 1999--2001.  There is no significant difference in the average
magnitude calculated from visual observations reported to VSNET between
the period after the outburst and the period of
1999--2001.\footnote{$\langle$http://vsnet.kusastro.kyoto-u.ac.jp/vsnet/index.html$\rangle$}

In figure \ref{fig:quies}, we also show three-night observations just
before the May outburst with the filled crosses (JD 2452397, 2452406,
and 2452407).  Their brightness is consistent with the ordinary
ellipsoidal modulation within the errors.  This means that neither
significant brightening nor fading was detected at least 13 d before the
outburst.  Visual observations reported to VSNET show the object was
near the quiescent level at least 2 d before the outburst.

\section{Discussion}
\subsection{Non-thermal optical emission during the outburst}

\citet{uem02v4641letter} suggest the presence of non-thermal optical
emission during the outburst of V4641 Sgr in May 2002 based on short
time scales of variations and unusual colors.  The optical SEDs around
the outburst maximum (JD 2452419) are shown in figure
\ref{fig:0524sed}.  As can be seen in this figure, during the bin
No. 7--11, the short-term variations became more prominent in the
lightcurve and the flat spectral component became more dominant in the
SEDs.  As proposed in \citet{uem02v4641letter}, the $\sim 100\,$s
variations definitely originated from the inner accretion region, where
we cannot expect strong optical thermal emission.  We conclude that the
flat spectral component was caused by a strong contribution of the
non-thermal emission, which was the source of short-term variations.

On the other hand, the object became bluer with the increase of the flux
during the large flare.  This property strongly indicates that the
flaring component was thermal emission.  The non-thermal short-term
variations apparently became stronger with the decrease of the thermal
flare. 

Considering the significant contribution of the non-thermal emission
even in $B$-band flux, we can estimate a lower limit of the temperature
of the thermal component by calculating with de-reddened $B$ and
$V$-band flux.  The temperature after the flare is estimated to be
typically $\gtrsim 14000\;{\rm K}$.  During this epoch, the contribution
of the non-thermal component is more than 15\% in $R_{\rm c}$-band and
30\% in $I_{\rm c}$-band.  On the other hand, at the peak of the flare
(bin No. 3 in figure \ref{fig:0524sed}), the temperature of the thermal
component is $\gtrsim 15000\;{\rm K}$.  The non-thermal contribution is
more than 10\% in $R_{\rm c}$-band and 25\% in $I_{\rm c}$-band.  It
should be noted that the non-thermal emission significantly contributes
to $R_{\rm c}$ and $I_{\rm c}$ bands even at the peak of the flare.

The non-thermal optical emission is furthermore supported by the flat
spectra in the radio range simultaneously taken with the optical
observation on JD 2452419.  In black hole binaries, flat or inverted
radio spectra have been interpreted as evidence for self-absorbed
synchrotron emission (\cite{fen01lowhardjet}).  Such synchrotron
emission has been proposed to possibly contribute to the optical emission
(\cite{fen01lowhardjet}; \cite{fen01j1118}; \cite{mar01j1118}).  In the
case of V4641 Sgr, based on i) the flat or inverted radio spectrum, ii)
the positive detection of polarization shown in figure
\ref{fig:0524radio}, and iii) the flat spectral component in the optical
energy spectra shown in figure \ref{fig:0524sed}, we conclude that,
around the outburst maximum, the wide range of the SED from the
radio--optical region was dominated by strong synchrotron emission.

As well as JD 2452419, a flat radio spectrum was also reported on JD
2452418.\footnote{$\langle$http://vsnet.kusastro.kyoto-u.ac.jp/vsnet/Mail/vsnet-campaign-v4641sgr/msg00037.html$\rangle$}.
On the other hand, the radio spectra just after the outburst were
optically-thin synchrotron emission as shown in figure 
\ref{fig:052526radio}.  The outburst may hence be characterized by
long-lived strong synchrotron emission or frequent synchrotron flares
which dominated in radio--optical wavelengths.  On the other hand, there
is also evidence for thermal flares, for example, the flare in figure
\ref{fig:0524sed} (and the flares just after the outburst, see the next
section).  It is possible that other multiple peaks during the outburst,
as seen in the panel (a) of figure \ref{fig:out}, were also a result of
thermal flares.  The complicated light curve of the May outburst is,
hence, probably due to the increasing activity of both thermal and
non-thermal sources, which must have an unknown relationship to each
other. 

\subsection{X-ray delay of flares just after the outburst}

In X-ray binaries, X-ray delays have been observed at onsets of X-ray
nova outbursts (\cite{oro97j1655precursor}; \cite{sha98v1333aql};
\cite{uem02j1118}).  They are interpreted as a result of the propagation
of a heating wave from an outer accretion disk to the inner disk.  The
X-ray nova outburst starts at the outer, low temperature portion of the
accretion disk, which first leads the optical brightening, and then, the
propagation of the hot region into the inner, high temperature portion
triggers the X-ray brightening later (\cite{oro97j1655precursor};
\cite{ham97j1655ADAF}).  This model has originally been developed for
the UV delay phenomenon observed in dwarf novae (\cite{liv92UVdelay};
\cite{mey94UVdelay}).  In the case of X-ray binaries, however, delay
times are so long (a few days) compared with those in dwarf novae (a few
hours) that another mechanism is required.  \citet{ham97j1655ADAF}
propose that the long delay times can be interpreted as a result of a
longer propagation time into the inner, an advection dominated accretion
flow (ADAF) region with a viscous time scale (\cite{nar94ADAF};
\cite{nar96BHXN}).  Based on the X-ray delay of order one day, the
transition radius from the standard disk to the ADAF is estimated to be
$\sim 10^4r_{\rm g}$ during the quiescent state.

The properties of the optical and X-ray variations on JD 2452420 are
summarized by i) the striking similarity of the light curves, ii) the
7-min X-ray delay, and iii) the shorter time scale of X-ray variations.
The X-ray delay phenomenon at the onset of X-ray nova outbursts also
have characteristics of i) and iii) (\cite{oro97j1655precursor}).  We
therefore conclude that the X-ray delay in V4641 Sgr can be also
understood in terms of a hot region propagating into the inner portion
of accretion flow.  This simple picture should, however, be modified to
explain the short time scale of the X-ray lag which is two orders
shorter than those observed in other X-ray binaries.

\citet{mar03v4641} reported that the X-ray spectrum on JD 2452420 can be
described not with a thermal disk, but with Compton scattering by a cold
neutral medium.  On the other hand, the source of the optical flares is
probably a hot, thermal accretion component in the outer disk.  We now
consider a situation that this hot region propagates into the inner
region with the viscous time scale.  Using the equation (5) of
\citet{ham97j1655ADAF} with same parameters (viscous parameter,
$\alpha=0.3$ and temperature, $T=10^4\;{\rm K}$), the 7-min X-ray lag
corresponds to the distance which the hot region travels, to be $\sim
r_{\rm g}$, where $r_{\rm g}$ is the Schwarzschild radius.  This small
distance means that the hot region propagates in the standard disk to
near the marginally stable orbit.  Under such situation, we can
generally expect strong thermal X-ray emission from the inner disk of
the temperature about $10^7\;{\rm K}$ (\cite{tan95BHXN}).  In
conjunction with the exceptionally low X-ray/optical flux ratio at the
flare peak ($L_{\rm X}/L_{\rm opt}\sim 1$), the observed
characteristics, that is, non-thermal and low luminosity X-ray emission,
are inconsistent with the simple picture. 

As well as the X-ray spectrum, the optical color is possibly
inconsistent with the simple picture.  Since the truncation radius is
quite small, we need to consider the propagation of the hot region in
the standard disk with the thermal time scale.  The propagation speed of
the hot region, $v_f$, can now be estimated with $v_f\sim \alpha c_{\rm
s}$, where $c_{\rm s}$ is the sound speed in the hot region
(\cite{mey84wave}).  Using the equation (1) in \citet{ham97j1655ADAF},
we obtain the source of the optical flare at $<5\times 10^8\;{\rm
cm}\; (=170r_{\rm g})$.  It is a much inner region compared with the
typical optical source at the outer accretion disk ($\sim
10^{10-11}\;{\rm cm}$).  On the other hand, the temperature of the hot
region can be estimated from the color.  The de-reddened color of
$B-V\sim 0.1$ indicates a hot region with the temperature of order 
$10^4\;{\rm K}$, which is the typical temperature at the outer accretion
disk during X-ray nova outbursts.   Since the temperature in a standard
disk strongly depends on the radius, it may hence be problematic that such
inner region of $170r_{\rm g}$ can produce flares of the low temperature
of $\sim 10^4\;{\rm K}$.

With these discussions, it is evident that the simple picture with a
propagating hot region must be improved to explain the observations.
The short X-ray delay may require a propagation time shorter than the
viscous or thermal time scale typical for black hole binaries.  If we
consider a significant contribution of non-thermal emission, as around
the outburst maximum, the temperature of the flare may actually be much
higher than $\sim 10^4$ K, which leads to a shorter thermal time scale.
Although the radio observation shows the significant fading and the
shift to the optically thin synchrotron emission 7 hours before the
flare (see, figures \ref{fig:0525x} and \ref{fig:052526radio}), it is
possible that a rapid radio flare occurred around the optical and X-ray
flares, as detected during the outburst.  It is also meaningful to
consider a situation where a hot region falls with the free-fall time
scale.  If this is the case, the optical emission source can locate at
outer region of $>10^{10}\;{\rm cm}$.  The ADAF can achieve such a
situation, however, the size of the ADAF would be much larger than those 
proposed in other black hole binaries.  An optically-thick advection
dominated flow, or a slim disk, can also reproduce the rapid
propagation.  \citet{rev02v4641} propose a similar situation to explain
the outburst of V4641 Sgr in 1999.  According to this model, the optical
emission originated from an intervening medium absorbing/reprocessing
X-rays, whose variation of size caused optical variations.  A similar
situation can explain the X-ray spectrum described with Compton
scattering, and possibly the optical and X-ray variations in figure
\ref{fig:0525x} and \ref{fig:0526x}.  In the case of the 2002 outburst
and the active phase, however, the luminosity of the object was so low
that we cannot expect the presence of a large, and long-lived
super-critical accretion, or slim disk.  

It is notable that the short X-ray delay was detected just after
the outburst, when the object was presumably in a transition phase from
the outburst to the quiescent state.  The X-ray lag on JD 2452421 was
longer than that on JD 2452420, which may imply that the object was just
in a rapid state transition to the ordinary quiescent phase. The rising
time scale of the flare was longer on JD 2452421 than that on JD
2452420.  If the similar flare observed on JD 2452430 (the panel (c) of
figure \ref{fig:active}) has the same nature, we can say that the rising
time scale was continuously increasing with time for about ten days.  If 
the rising time scale corresponds to the propagation time of the hot
region, the increase of the rising time scale may indicate the expansion
of the accretion disk.

\subsection{The peculiarity of the optical flash}

Although optical short-term modulations were detected during the
outburst in 2002 May, the optical flash on JD 2452462--2452463 has
noteworthy characteristics.  First, the object was not in a major
outburst, but remained at the quiescent level around JD 2452463.  During
the outburst in May, the object experienced repeated short flares and
the total duration was six days.  On the other hand, variations like in
figure \ref{fig:flash} were observed only in this period and not on JD
2452461 and 2452464, which indicates a short duration of this active
state.  Second, no optical flash was detected during the May outburst.
The $\sim 100$ s variations during the May outburst have amplitudes of
order 0.1 mag, while the optical flash has amplitudes of 1.2 mag.
The unique feature of the optical flash is the huge released energy only
within the time scale of 10 s.

Using the peak apparent magnitude of $R_{\rm c}=12.4$ mag and the
quiescent magnitude ($V=13.8$), the optical luminosities are estimated
to be $L_{\rm opt,peak}>5.2\times 10^{36}\;{\rm erg\, s^{-1}}$ and
$L_{\rm opt,quies}=1.6\times 10^{36}\;{\rm erg\, s^{-1}}$, respectively.  
In the above estimations, we assume an interstellar
extinction of $A_V\sim 1.0$ and a distance of $d\sim 9.6\;{\rm kpc}$,
which are reported in \citet{rev02v4641} and \citet{oro01v4641sgr}.  It
should be noted that our observations with exposure times of $5$ s may
have overlooked more rapid variations and the real peak of the flash,
and hence can only provide a lower limit of $L_{\rm opt,peak}$ in the
above estimation.  The peak released-energy rate of the flash component
is then calculated to be $L_{\rm flash} > L_{\rm opt,flash} = L_{\rm
opt,peak}-L_{\rm opt,quies}\gtrsim 4\times 10^{36}\;{\rm erg\, s^{-1}}$,
where $L_{\rm opt,flash}$ is the observed optical luminosity of the flash.

On the other hand, we can estimate a theoretical upper limit of the
energy-release rate at a certain portion of an accretion disk.  The
observed optical emission on JD 2452463 indicates that the peak
luminosity of the flash was much lower than that of the peak of its
super-Eddington outburst in 1999 September (\cite{uem02v4641};
\cite{rev02v4641}).  The mass accretion rate was hence definitely
smaller than the critical accretion rate ($\dot{M}_{\rm crit}\equiv
L_{Edd}/\eta c^2$) throughout the active phase, where $\eta$ is the
energy conversion rate.  We set a secure upper limit of the mass
accretion rate to be $<0.1\dot{M}_{\rm crit}$ based on the optical peak
luminosity of the flash.  Under these assumptions, the observed energy
release rate needs an emission source at
\begin{eqnarray}
R_{\rm flash} <150({\eta\over 0.1})r_{\rm g}.
\end{eqnarray}
The radius of $150r_{\rm g}$ corresponds to a dynamical time scale of 1
s.  It is then possible that the optical flux was variable at 1-s scale
during the active phase.  Our visual observations indeed detected a very
rapid variations of order 1 s and amplitudes of 0.3--0.9 mag
during the active phase in 2002
May--July.\footnote{$\langle$http://vsnet.kusastro.kyoto-u.ac.jp/vsnet/Mail/vsnet-campaign-v4641sgr/msg00008.html,msg00026.html, msg00151.html $\rangle$}
These observations may support the optical
emission source at $<150r_{\rm g}$.

Such an inner emission source strongly indicates that the optical
emission was non-thermal (\cite{uem02v4641letter}).  The optical flashes
have a time scale shorter than that observed during the outburst in
2002 May, and hence, the scenario with synchrotron emission is even more
preferable.  The energy conversion rate in equation (1) strongly depends
on the mechanism of the optical flash.  When we consider a synchrotron
optical flash, a part of gravitational energy is first transformed to
magnetic energy, which then leads the acceleration of particles and
generates synchrotron emission.  The energy conversion rate should
therefore be much smaller than 0.1 which is the value for the most
effective disk.  If we consider $\eta\sim 0.01$--$0.001$, it is possible
that the optical flash is an event which occurred at the innermost
region of the accretion flow.

When we consider the mechanism of the flash, it is interesting to note
that the dips just before the flashes are also reported in the optical
short-term variations in XTE J1118+480 (\cite{kan01nature};
\cite{spr02j1118}).  Both variations have a time scale of a few seconds,
however, we should note that the released energy during the optical
flash is much greater than those of rapid optical variations in XTE
J1118+480 and GX 339$-$4 (\cite{ste97gx339}; \cite{spr02j1118}).

\section{Summary}

We observed V4641 Sgr during the outburst and post-outburst active phase
in 2002 at optical, radio, and X-ray wavelengths.  In the optical range,
the object exhibited strong variations with a wide range of periods
of $10^2$--$10^4$ s.  We conclude that $\sim 1000$ s, $>1\;{\rm
mag}$ flares are thermal emission based on their bluer color and the
X-ray lag.  The 7-min X-ray lag is, however, too small to be interpreted
in terms of a hot region propagating into the inner disk, as in the
onset of X-ray nova outbursts.  The short X-ray lag requires an unknown
mechanism to account for the short propagation time of the hot region.
On the other hand, the origin of short-term, $\sim 100\;{\rm s}$
variations was presumably synchrotron emission.  The flattening of the
optical spectrum at the $I_{\rm c}$, $R_{\rm c}$, and $V$-band region
strongly indicates a significant contribution of synchrotron emission
even in the optical range.  Their amplitudes are even relatively large
(0.1--0.5 mag) compared with other black hole binaries showing optical
rapid variations.  Contrary to the current framework of the optical
rapid variations, the optical flash, which was detected only on JD
2452462--2452463, has the unique and surprising characteristics that the
object brightened by 1.2 mag only within 30 s.  The released energy
indicates that the flash occurred at the innermost region of the accretion
flow.  It should be noticed that dips were observed just before the
flashes, as in XTE J1118+480. 

To date, V4641 Sgr has continued to show us different features whenever
it entered active phases.  Considering the short duration of an active
phase, we have possibly overlooked a number of other active phases.
V4641 Sgr has almost all topics to which we have recently paid
attention, for example, the disk--jet relationship, the broad Fe
emission line in the X-ray range, the non-thermal optical emission, and
the rapid optical variation.  Therefore, close monitoring in all
wavelength is strongly urged not only to reveal the nature of V4641 Sgr,
but also to fully understand the accretion physics around the black
hole.

\vskip 3mm

We are grateful to many amateur observers for supplying their vital
visual and CCD estimates via Variable Star Network (VSNET;
http://vsnet.kusastro.kyoto-u.ac.jp/vsnet/).  This work is partly
supported by a grant-in aid from the Japanese Ministry of Education,
Culture, Sports, Science and Technology (No. 13640239, 15037205).  
Part of this work is supported by a Research Fellowship of the Japan
Society for the Promotion of Science for Young Scientists (MU and RI).

\end{document}